# Target-Dependent Chemical Species Tomography with Hybrid Meshing of Sensing Regions

Rui Zhang, Jingjing Si, *Member, IEEE*, Godwin Enemali, *Member, IEEE*, Yong Bao, Chang Liu, *Member, IEEE*

*Abstract*—This paper develops a hybrid-size meshing scheme for target-dependent imaging in Chemical Species Tomography (CST). The traditional implementation of CST generally places the target field in the central region of laser sensing, the so-called Region of Interest (RoI), with uniform-size meshes. The centre of the RoI locates at the midpoint between the laser emitters and receivers, while the size of the RoI is empirically determined by the optical layout. A too small RoI cannot make the most use of laser beams, while a too large one leads to much severer rank deficiency in CST. To solve the above-mentioned issues, we introduce hybrid-size meshing, for the first time, by reforming the density of the pixels in the entire sensing region of CST. This development alleviates the ill-posedness of the CST inverse problem by detailing the target flow field with dense pixels in the RoI and fully considering the complete physical absorption model with sparse pixels out of the RoI. The proposed scheme was both numerically and experimentally validated using a CST sensor with 32 laser beams using a variety of computational tomographic algorithms. The images reconstructed using the hybrid-size meshing scheme show better accuracy and finer profile of the target flow, compared with those reconstructed using the traditionally uniform-size meshing. The proposed hybrid-size meshing scheme significantly facilitates the industrial application of CST towards practical combustors, in which the combustion zone is bypassed by cooling air. In these scenarios, the proposed scheme can better characterise the combustion zone with dense meshes, while maintaining the integrity of the physical model by considering the absorption in the bypass air with sparse meshes.

## I. Introduction

IN recent years, Chemical Species Tomography (CST) has become one of the most representative techniques used in combustion diagnosis for rapid imaging of unknown two-dimensional (2D) distributions of flow parameters, such as species concentration and temperature [1-4]. CST is implemented by multiple line-of-sight tunable diode laser absorption spectroscopy (TDLAS) measurement, in a manner analogous to x-ray tomography. Its robustness and minimal intrusiveness enable CST a highly desired and *in situ* solution for industrial application, e.g. vapour fuel imaging in internal combustion engines [5, 6] aero-engine lean blowout diagnosis

[7] and gas turbine exhaust imaging [8, 9].

CST is typically implemented by placing the target field in the central region of laser sensing, the so-called Region of Interest (RoI), in which the distributions of the target parameters are retrieved using tomographic algorithms. The RoI, typically with its centre located at the midpoint between the laser emitters and receivers, is mostly segmented with uniform-size meshes. The size of the RoI can be chosen to (a) cover the whole sensing region [10, 11] or (b) the central part of the sensing region that mainly covers the target field [12-14]. However, both cases suffer from defects illustrated as follows:

Case (a): Due to the limited optical access in industrial practice, CST is inherently and inevitably ill-posed with severe rank deficiency. Given an RoI that covers the entire sensing region, some areas within the RoI have few or even no access to the laser beams [11]. With uniform-size meshes, beam layout with limited projections worsens the ill-posedness of the CST inverse problem, resulting in significant spike noise and artefacts in image reconstruction.

Case (b): This solution, to some extent, mitigates the ill-posedness by aggregating the sensing capability of CST in the RoI. Previous attempts used Nitrogen to purify the area outside the RoI [12], or assumed the small absorption outside can be neglected [13]. However, the former is less practical to be deployed on industrial combustors, while the latter suffers from errors in the reconstructions due to the physical existence of outward-RoI heat dissipation and species convection.

From the mathematical perspective, regularisation methods, such as Tikhonov regularisation and truncated singular value decomposition (TSVD) [15], can improve the robustness of image reconstruction at a proper cost of accuracy. However, they have no contribution to spatial sampling and thus is ineffective to complement any physical information of the tomographic reconstruction. In addition, some efforts were made from the perspectives of modifying the sensors and beam arrangements, for example introducing both parallel- and fan-beam projections [16]. However, the limited optical access on practical combustors largely invalids these attempts. Based on the existing beam arrangement, it is highly demanded to develop a limited-data tomographic strategy that can both detail the reconstruction of target field and fully consider the physical

This work was supported in part by the UK Engineering and Physical Sciences Research Council under Grants EP/P001661/1 and EP/T012595/1. *(Corresponding author: Chang Liu.)*

R. Zhang, G. Enemali, Yong Bao, and C. Liu are with the School of Engineering, University of Edinburgh, Edinburgh EH9 3JL, U.K. (e-mail: C.Liu@ed.ac.uk).

existence of the absorption in the sensing region.

To address the above-mentioned requirements, we introduce target-dependent CST with hybrid-size meshes in the entire laser sensing region. Although a couple of previous works reported CST in irregular sensing regions, these attempts are still based on a uniform-size or nearly uniform-size meshing schemes, and have not customized the density of the grid cell towards the target flow to be reconstructed. For example, Wood et al. [17] numerically demonstrated the reconstruction of an annular measurement domain for turbofan engines imaging, but with uniform-size meshes in the domain. Recently, Grauer et al. [18] recently applied the finite element in meshing with Bayesian model for image reconstruction in irregular domains. The finite element meshing scheme allows more flexibility at the domain boundary. However, the meshes in the domains are still nearly uniform-size without specific or customized discretization of the target area. In our work, dense meshes are allocated in the RoI to detail the target flow field and sparse ones out of the RoI to consider the physically existing molecular absorption. This new meshing scheme significantly facilitates the industrial application of CST towards practical combustors without modifying the layout of the optical sensor. In case of imaging combustion zone that is bypassed by cooling air, e.g. aero-engine exhaust imaging, the proposed scheme can better characterise the target combustion zone with dense meshes, while maintaining the integrity of the physical model by considering the absorption in the bypass air with sparse meshes.

The rest of the paper is structured as follows: The proposed scheme is mathematically analysed in Section II to address its improvement on the tomographic image quality. Numerical simulation and experiments are carried out to validate the proposed meshing scheme in Section III and Section IV, respectively. The paper is concluded in Section V.

## II. METHODS

### A. Mathematical Formulation of CST

The fundamentals of CST are firstly reviewed to facilitate the introduction of the hybrid-size meshing scheme in the next subsection. As shown in Fig. 1, the laser sensing area that covers entire optical paths from emitters to detectors is named as Region of Sensing (RoS) here afterwards.

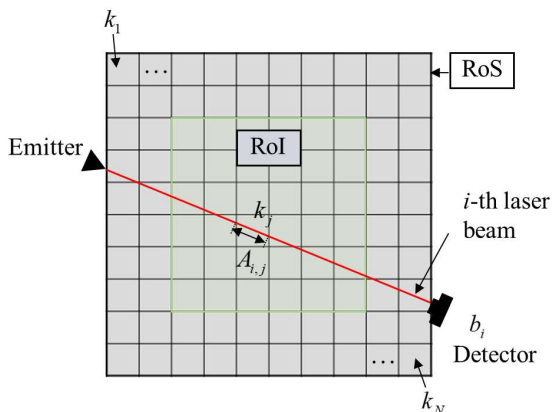

Fig. 1. Geometric description of an exampled laser beam in CST.

Given the RoS is discretised into $N$ pixels, the temperature, pressure, and mole fraction (concentration) of the absorbing species are assumed to be uniformed in each pixel. According to Beer's law, the integrated absorbance $b_i$ of the $i$-th laser beam for the target species is given by

$$b_i = \sum_{j=1}^{N} A_{i,j} k_j, \qquad (1)$$

where $A_{i,j}$ represents the chord length of the $i$-th beam within the $j$-th pixel, $k_j$ the absorption density of the $j$-th pixel at the selected absorption transition [2]. $k_j$ is defined by

$$k_j = P_j x_j S(T_j), \qquad (2)$$

where $P_j$ [atm] is the pressure, $x_j$ the mole fraction of the absorbing species, $T_j$ [K] the temperature, $S(T_j)$ [cm⁻²·atm⁻¹] the line strength of the transition in the $j$-th pixel, respectively.

For a tomographic system with $M$ laser beams, (1) can be formulated as a linear equation

$$\mathbf{Ak} = \mathbf{b} = \mathbf{b}_{in} + \mathbf{b}_{out}, \qquad (3)$$

where $\mathbf{A} \in \mathbb{R}^{M \times N}$ is the sensing matrix, $\mathbf{k} \in \mathbb{R}^{N}$ the vector of pixel-wised absorption density $k_j$ ($j=1, 2, …, N$) to be solved in the inverse problem. $\mathbf{b} \in \mathbb{R}^{M}$ is the vector of the measured integrated absorbance $b_i$ ($i=1, 2, …, M$), which is a summation of integrated absorption in the RoI, $\mathbf{b}_{in} \in \mathbb{R}^{M}$ and integrated absorption out of the RoI $\mathbf{b}_{out} \in \mathbb{R}^{M}$.

If the region out of the RoI is not Nitrogen purified, the most common case in industrial applications, $\mathbf{b}_{in}$ and $\mathbf{b}_{out}$ cannot be separated before the reconstruction. Taking only the RoI as the reconstruction region without properly considering $\mathbf{b}_{out}$ will definitely lead to reconstruction error in the tomographic image.

### B. Hybrid-size RoS meshing scheme

As illustrated in [19] and [20], the inverse problem described in (3) is inherently ill-posed that can be illustrated through Singular Value Decomposition (SVD) of the sensing matrix $\mathbf{A}$:

$$\mathbf{A} = \mathbf{USV}^{\mathrm{T}}, \qquad (4)$$

where $\mathbf{U} \in \mathbb{R}^{M \times M}$ and $\mathbf{V} \in \mathbb{R}^{N \times N}$ are orthonormal matrices, and $\mathbf{S} \in \mathbb{R}^{M \times N}$ a diagonal matrix containing the singular values in descending order.

In principle, the inverse problem aims at finding a unique solution, $\mathbf{k}^{LS} \in \mathbb{R}^{N}$ to minimise the least square error, that is, $\mathbf{k}^{LS} = \arg\min\left(\left\|\mathbf{Ak} - \mathbf{b}\right\|^2\right)$. When $M > N$, this solution can be calculated by

$$\mathbf{k}^{LS} = \sum_{j=1}^{N} \frac{\mathbf{u}_j^{\mathrm{T}} \mathbf{b}}{\sigma_j} \mathbf{v}_j, \qquad (5)$$

where $\mathbf{u}_j$ and $\mathbf{v}_j$ are the $j$-th column vectors of $\mathbf{U}$ and $\mathbf{V}$, respectively. $\sigma_j$ is the $j$-th singular value in the diagonal $\mathbf{S}$.

In practice, the number of laser beams $M$ is limited by optical access to the combustors, resulting into $M < N$. In this case, $\mathbf{k}^{LS}$ in (5) should be separated into two parts described in (6): the unique solution $\mathbf{k}^{unique} \in \mathbb{R}^{N}$ that gives the minimised value of $\|\mathbf{Ak\text{-}b}\|^2$ satisfying $\mathbf{Ak} = \mathbf{b}$, and the non-unique solution $\mathbf{k}^{\text{null}} \in \mathbb{R}^{N\text{-}M}$ from solving $\mathbf{Ak} = \mathbf{0}$.

$$\mathbf{k}^{\text{LS}} = \mathbf{k}^{\text{unique}} + \mathbf{k}^{\text{null}} \qquad (6)$$

Considering the practical measurements $\mathbf{b}$ is a superposition of noise-free data, $\mathbf{b}^{\text{true}}$, and the noise, $\mathbf{b}^{\text{noise}}$. The composition of $\mathbf{k}^{\text{unique}}$ can be expressed by

$$\mathbf{k}^{\text{unique}} = \sum_{j=1}^{M} \frac{\mathbf{u}_j^{\text{T}}\mathbf{b}^{\text{true}}}{\sigma_j}\mathbf{v}_j + \sum_{j=1}^{M} \frac{\mathbf{u}_j^{\text{T}}\mathbf{b}^{\text{noise}}}{\sigma_j}\mathbf{v}_j \qquad (7)$$

Since $\mathbf{A}$ is rank-deficient, the lack of $N - M$ measurements causes nontrivial null-space in $\mathbf{A}$, thus leading to $N - M$ undetermined values in $\mathbf{k}^{\text{null}}$. Hence, (6) can be further expanded as:

$$\mathbf{k}^{\text{LS}} = \sum_{j=1}^{M} \frac{\mathbf{u}_j^{\text{T}}\mathbf{b}^{\text{true}}}{\sigma_j}\mathbf{v}_j + \sum_{j=1}^{M} \frac{\mathbf{u}_j^{\text{T}}\mathbf{b}^{\text{noise}}}{\sigma_j}\mathbf{v}_j + \sum_{j=M+1}^{N} c_j\mathbf{v}_j, \qquad (8)$$

where $\mathbf{c} \in \mathbb{R}^{N\text{-}M}$, with the $j$-th element $c_j$, is a set of undetermined scalars to describe $\mathbf{k}^{\text{null}}$.

The ill-posedness of (8) can be described in two aspects:

(a) Noise susceptible. As indicated by the second term in (8), $\mathbf{b}^{\text{noise}}$ can be significantly magnified with relatively small singular values $\sigma_j$, worsening the quality of the reconstructed images.

(b) Solution underdetermined. In the case of $M < N$, the singular value does not exist for $j > M$. The third term in (8) is the indicator of undetermined solutions. The larger the $N - M$ is, the more severely undetermined solution will exist.

To alleviate the ill-posed problem mentioned above, we propose the hybrid-size meshing scheme to address the target flow in the RoS. With hypothesis that the RoI is more important than the region out of the RoI, the new scheme (a) mitigates the impact of noise by refactoring the sensing matrix and (b) improves spatial resolution in the RoI by reallocate the size of meshes. An example gives in Fig. 2. The optical layout has a total of 32 laser beams arranged with 4 equiangular projection and 8 equispaced parallel beams in each projection. Given the RoS with dimensions of $L \times L$, the central RoI is defined by $L/2$

$\times L/2$. As shown in Fig. 2 (a), more laser beams penetrate the pixels in the RoI. In contrast, the pixels out of the RoI are partially penetrated by several beams. It is worth noting that small numbers of pixels and laser beams are used in the figure for a clear view of the meshes and beam arrangement. The proposed scheme is also suitable for practical application with more densely arranged laser beams and finer discretised sensing regions.

For the hybrid-size meshing scheme, shown in Fig. 2 (b), dense pixels with dimensions of $L/10 \times L/10$ are deployed in the RoI to achieve better spatially resolved reconstruction of the target flow. Sparse pixels with dimensions of $L/5 \times L/5$ are deployed out of the RoI to fully consider the physically existing molecular absorption introduced by heat radiation and species convection. As a result, the entire RoS is discretised into 52 pixels, with 36 pixels in the RoI, noted as $P_{\text{RoI}}^{\text{H}}$, and 16 pixels out of the RoI.

In comparison with the hybrid-size meshes, two RoS with uniform-size meshes are given in Fig. 3. To achieve a similar number of the pixels as the hybrid-size meshing, the uniform-size meshing scheme discretises the RoS in Fig. 3 (a) into 49 pixels with dimensions of $L/7 \times L/7$, from which 25 pixels are in the central RoI, noted as $P_{\text{RoI}}^{\text{U1}}$. In addition, the entire RoS in Fig. 3 (b) has 100 uniform-size pixels with dimensions of $L/10 \times L/10$, resulting into same resolution and pixel number in the RoI as those in the hybrid-size meshing, i.e. $P_{\text{RoI}}^{\text{U2}} = P_{\text{RoI}}^{\text{H}} = 36$.

Theoretically, the hybrid-size meshing scheme modifies the structure of the sensing matrix $\mathbf{A}$. Given $M$ laser beams, the sensing matrix for the hybrid-size meshing scheme, note $\mathbf{A}_{\text{Hybrid}}$, is obtained by concatenating the sensing matrix in the RoI, $\mathbf{A}^{\text{in}} \in \mathbb{R}^{M \times P_{\text{RoI}}^{\text{H}}}$, with that out of the RoI, $\mathbf{A}^{\text{out}} \in \mathbb{R}^{M \times (N - P_{\text{RoI}}^{\text{H}})}$.

$$\mathbf{A}_{\text{Hybrid}} = \begin{bmatrix} a_{1,1}^{\text{in}} & \cdots & a_{1,P_{\text{RoI}}^{\text{H}}}^{\text{in}} & a_{1,(P_{\text{RoI}}^{\text{H}}+1)}^{\text{out}} & \cdots & a_{1,N}^{\text{out}} \\ \vdots & \ddots & \vdots & \vdots & \ddots & \vdots \\ a_{M,1}^{\text{in}} & \cdots & a_{M,P_{\text{RoI}}^{\text{H}}}^{\text{in}} & a_{M,(P_{\text{RoI}}^{\text{H}}+1)}^{\text{out}} & \cdots & a_{M,N}^{\text{out}} \end{bmatrix}, \qquad (9)$$

$$\underbrace{\phantom{a_{1,1}^{\text{in}} \cdots a_{1,P}^{\text{in}}}}_{\mathbf{A}^{\text{in}} \in \mathbb{R}^{M \times P_{\text{RoI}}^{\text{H}}}} \underbrace{\phantom{a^{\text{out}} \cdots a^{\text{out}}}}_{\mathbf{A}^{\text{out}} \in \mathbb{R}^{M \times (N - P_{\text{RoI}}^{\text{H}})}}$$

where $a^{\text{in}}$ and $a^{\text{out}}$ are the elements in $\mathbf{A}^{\text{in}}$ and $\mathbf{A}^{\text{out}}$, denoting the chord lengths of the laser path in a pixel in and out of the RoI, respectively. Given $P_{\text{RoI}}^{\text{H}} = 36$ for the hybrid-size meshes shown in Fig. 2 (b), the numbers of columns in $\mathbf{A}^{\text{in}}$ and $\mathbf{A}^{\text{out}}$ are

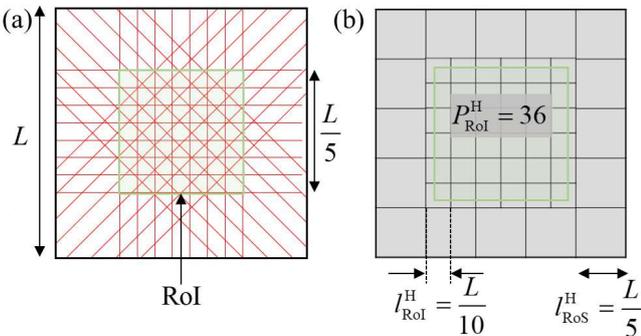

Fig. 2. Demonstration of the hybrid-size meshing scheme. (a) shows an example optical layout with 4 equiangular projection and 8 equispaced parallel beams in each projection. (b) shows the hybrid-size meshes with a total of 52 pixels, 36 pixels in the RoI and 16 pixels out of the RoI.

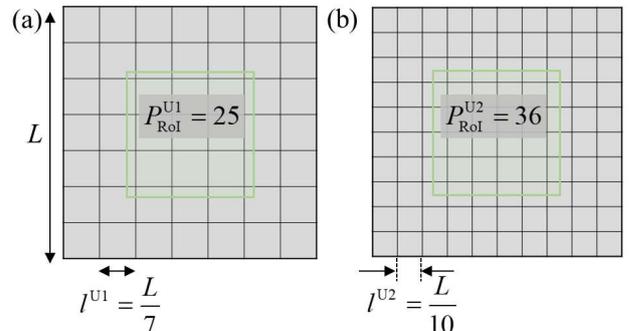

Fig. 3. Two Regions of Sensing (RoS) discretised using the uniform-size meshing scheme. (a) and (b) have 49 and 100 pixels with dimensions of $L/7 \times L/7$ and $L/10 \times L/10$, respectively.

36 and 16, respectively.

Similarly, the sensing matrix of the uniform-size meshing scheme, note $\mathbf{A}_{7\times7}$, can be described as:

$$\mathbf{A}_{7\times7} = \underbrace{\begin{bmatrix} a_{1,1}^{in} & \cdots & a_{1,P_{RoI}^{U}}^{in} \\ \vdots & \ddots & \vdots \\ a_{M,1}^{in} & \cdots & a_{M,P_{RoI}^{U}}^{in} \end{bmatrix}}_{\mathbf{A}^{in}\in\mathbb{R}^{M\times P_{RoI}^{U}}} \underbrace{\begin{bmatrix} a_{1,(P_{RoI}^{U}+1)}^{out} & \cdots & a_{1,N}^{out} \\ \vdots & \ddots & \vdots \\ a_{M,(P_{RoI}^{U}+1)}^{out} & \cdots & a_{M,N}^{out} \end{bmatrix}}_{\mathbf{A}^{out}\in\mathbb{R}^{M\times(N-P_{RoI}^{U})}}, \quad (10)$$

where $\mathbf{A}^{in} \in \mathbb{R}^{M \times P_{RoI}^{U}}$ and $\mathbf{A}^{out} \in \mathbb{R}^{M \times (N-P_{RoI}^{U})}$ are sensitivity matrices in and out of the RoI for the uniform-size meshing scheme, respectively. Given $P_{RoI}^{U1} = 25$, the numbers of columns in $\mathbf{A}^{in}$ and $\mathbf{A}^{out}$ for the uniform-size meshes shown in Fig. 3 (a) are 25 and 24, respectively.

TABLE I statistically demonstrates the properties of the two example sensing matrices, $\mathbf{A}_{Hybrid}$ and $\mathbf{A}_{7\times7}$. The table includes the dimension of each sensing matrix, number of columns for $\mathbf{A}^{in}$, number of columns for $\mathbf{A}^{out}$, and percentage of non-zero values in each matrix. It can be seen the hybrid-size meshing leads to a larger proportion of pixels in the RoI than the uniform-size meshing. This is indicated by the larger number of columns for $\mathbf{A}^{in}$ in $\mathbf{A}_{Hybrid}$ than that in $\mathbf{A}_{7\times7}$. As a result, the hybrid-size meshing not only improve the spatial resolution of the RoI, but also dedicates more computation resource to reconstruct the RoI. In addition, the hybrid-size meshing gives the higher percentage of non-zero values, indicating its capability of reducing the number of pixels without beam passing through.

The performance of image reconstruction using the two meshing schemes can be quantified by plotting the singular values of the sensitivity matrices. For simplicity, the last nontrivial rows in both matrices obtained from the two meshing schemes were duplicated and extended into the dimensions of $N \times N$. $N$ equals to 52 and 49 for the hybrid-size and uniform-size meshes, respectively. As a result, both yield $N$ singular values in the diagonal $\mathbf{S}$. Fig. 4 shows the singular values $\sigma_j$ ($j$ =1, 2, ..., $N$) in descending order obtained from each meshing scheme with the undetermined $\sigma_j$ equalling to zero. Given $j >$ 18, $\sigma_j$ obtained using the hybrid-size meshes are not only larger than those obtained using the uniform-size meshes but also perform a slower decay to zero. As indicated by the second term of (8), the reduced number of extremely small singular values will impose stronger suppression to the measurement noise, thus contributing to better image quality in CST.

TABLE I
PROPERTIES OF THE SENSING MATRICES OBTAINED USING THE HYBRID-SIZE AND UNIFORM-SIZE MESHING SCHEMES

|  | Size | Columns in $\mathbf{A}^{in}$ | Columns in $\mathbf{A}^{out}$ | Percentage of non-zero values |
|---|---|---|---|---|
| $\mathbf{A}_{Hybrid}$ | $\mathbb{R}^{32\times52}$ | 36 | 16 | 18.75% |
| $\mathbf{A}_{7\times7}$ | $\mathbb{R}^{32\times49}$ | 25 | 25 | 17.35% |

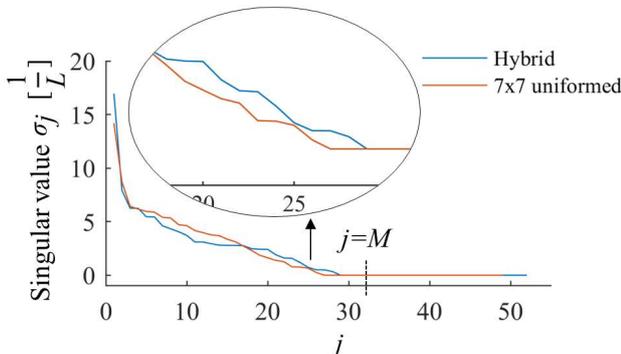

Fig. 4. Comparison between the singular values of the sensitivity matrices obtained using the hybrid-size meshing scheme with 52 pixels and the uniform-size meshing scheme with 49 pixels.

To achieve the same pixel size in the RoI as the hybrid-size meshes, the uniform-size meshes in Fig. 3 (b) overweight the solutions out of the RoI. Given $P_{RoI}^{U2} = 36$ for the uniform-size meshes shown in Fig. 3 (b), the numbers of columns in $\mathbf{A}^{in}$ and $\mathbf{A}^{out}$ are 36 and 64, respectively. In other words, there are 64 undetermined solutions when adopting the uniform-size meshes in Fig. 3 (b). However, the number of undetermined solutions achieved by the hybrid-size mesh is only 16. As indicated by the third term of (8), the smaller number of undetermined solutions improves the performance of image reconstruction by reducing the uncertainty of $\mathbf{k}^{LS}$ retrieval.

### C. Image reconstruction algorithm

As the proposed meshing scheme essentially reshapes the structure of the ill-conditioned sensing matrix $\mathbf{A}$ defined in (3), it can be widely suitable for the computational tomographic algorithms that are dependent on solving the minimization problem in (11).

$$\arg\min\left\{\left\|\mathbf{b} - \mathbf{Ak}\right\|_2^2\right\}. \quad (11)$$

Here, three computational tomographic algorithms that satisfies (11), i.e. Tikhonov regularization (TK) [21], algebraic reconstruction technique (ART) [22, 23], and total variation (TV) [24], are employed to validate the wide adaptability of hybrid-size meshing scheme, with their performances compared subsequently.

Considering the smooth distribution of k, TK solves the inverse problem by:

$$\arg\min\left\{\left\|\mathbf{b} - \mathbf{Ak}\right\|_2^2 + \gamma\left\|\mathbf{Fk}\right\|_2^2\right\}, \text{ s.t. } \mathbf{k} \geq 0, \quad (12)$$

where $\|\mathbf{Fk}\|_2^2$ is the first-order Tikhonov regularisation term with a linear differential operator $\mathbf{F}$, $\gamma$ the empirically determined regularisation parameter.

As one of the most representative iterative tomographic algorithms in CST, ART solves the linear equation set (3) iteratively by

$$\mathbf{k}^{(z+1,i)} = \mathbf{k}^{(z,i)} + \lambda \frac{\mathbf{A}_i\mathbf{k}^{(z,i)} - b_i}{\left\|\mathbf{A}_i\right\|_2^2}\mathbf{A}_i^{\mathrm{T}}, \quad (13)$$

where $z$ is the index of iterations. For each iteration, $i$ enumerates all beams one by one. $\mathbf{A}_i$ denotes the $i$-th row of $\mathbf{A}$. $\lambda$ is the relaxation coefficient.

TV algorithm solves (3) by minimization of the l1-norm of the image total variation as follows:

$$TV(k) = \underset{m,n}{\operatorname{argmin}}(\sum_{m,n}\sqrt{(k_{m+1,n}-k_{m,n})^2+(k_{m,n+1}-k_{m,n})^2})\,\text{s.t.}\,\mathbf{A}k = \mathbf{b},$$

$$(14)$$

where $m$ and $n$ are the row and column indices of the reconstructed image, $k_{m,n}$ refers to the element in $\mathbf{k}$ that contains the results at $(m, n)$ pixel in the reconstruction image.

With $\mathbf{k}$ in hand, the species concentration for each pixel, $x_j$, is calculated from (2).

## III. NUMERICAL VALIDATION

Numerical simulation was carried out to validate the proposed hybrid-size meshing scheme by reconstruction of two sets of phantoms with ART. The reconstructed images were then compared with those obtained using the uniform-size meshing scheme with quantified image errors.

### A. Simulation setup

Water vapour ($H_2O$) is one of the major products of hydrocarbon combustion and its distribution is interested in industrial combustion community for the evaluation of combustion efficiency. In this paper, the $H_2O$ absorption transition centred at $v = 7185.6$ cm$^{-1}$ was selected to reconstruct the distributions of $H_2O$ concentration since it has appropriate linestrength to give a good signal to noise ratio (SNR) for the TDLAS measurement [25].

As a cost-effective optical layout for CST, parallel beam arrangement has been widely adopted for the tomographic sensor design in practical applications [12, 26]. As shown in Fig. 5, a parallel beam arrangement was used in this work with 32 laser beams arranged in 4 equiangular projection angles, each angle with 8 equispaced parallel beams. The angular spacing between each projection angle is 45°, while the beam spacing within a projection angle, noted as $d$, is 1.8 cm. The distance between each pair of laser emitter and detector, noted as $D$, is 36.8 cm, enclosing the RoS with an octagonal shape. The RoI is defined as the central square region with side length $l_{RoI} = 12.8$ cm.

The practical distributions of $H_2O$ concentration in diffusion flows were simulated using Large Eddy Simulation (LES) via Fire Dynamic Simulator (FDS) [27, 28]. The simulation domain is a $36.8 \times 36.8 \times 10$ cm$^3$ rectangular space with top and four side boundaries opened, which is divided into $280 \times 280 \times 20$ cells. The domain was filled with air. Its temperature and $H_2O$ mole fraction were considered as uniform with 294.15 K and 0.005, respectively. To generate the inhomogeneous $H_2O$ distributions, water vapour was jet from a circular inlet at the bottom of the domain with a constant velocity. The cross section for CST measurement was set at 1 cm above the inlet. In this work, two scenarios were considered, one with the inlet located at the centre, the other one with the inlet located to the bottom right of the centre. Table II details the inlet radius, $r$ [cm], central location of the inlet, $(x, y)$ [cm], and the jet velocity, $v$ [m/s] for each scenario. To obtain high-accuracy path integrated absorbances $\mathbf{b}$ in the forward problem in (3), two sets of high-resolution phantoms were generated with 10136 pixels in the RoS, each with 0.13 cm × 0.13 cm. Starting by jetting the water vapour from the inlet, Media 1 and Media 2 record the dynamic true distributions of $H_2O$ concentration for scenario 1 and scenario 2, respectively. For each scenario, a total of 50 frames, sampled at an interval of 0.2 s, were processed to make the video shown in this work. Two instantaneous distributions of $H_2O$ concentration, i.e. 15th frames in Media 1 and Media 2, are shown in Fig. 6.

### B. Simulation results and discussion

In the simulation, the phantoms shown in Fig. 6 were reconstructed using both the proposed hybrid-size meshing scheme and the uniform-size meshing scheme. For the hybrid-size meshing, the dimensions of the pixels in and out of the RoI were chosen to be 1.84 cm × 1.84 cm and 3.68 cm × 3.68 cm, respectively. As shown in Fig. 7 (a), the RoS is segmented into 196 pixels, among which 144 pixels are in the RoI. To obtain a similar number of pixels in the RoS for both meshing scheme, the dimension of the pixels using the uniform-size meshing scheme shown in Fig. 7 (b) is 2.63 cm × 2.63 cm, resulting in 172 pixels in the RoS with 64 pixels in the RoI.

Since the regularisation parameter $\gamma$ in (12) plays an important role in image reconstruction, the optimal $\gamma$ was selected for each phantom based on quantification of Image Error (IE) in the RoS. In this work, IE is defined as the normalised root mean square error between the reconstructed

TABLE II
SIMULATION PARAMETERS OF THE TWO SCENARIOS.

| | $r$ [cm] | $(x, y)$ [cm] | $v$ [m/s] |
|---|---|---|---|
| *Scenario 1* | 5 | (0, 0) | 0.25 |
| *Scenario 2* | 4 | (3, -3) | 0.25 |

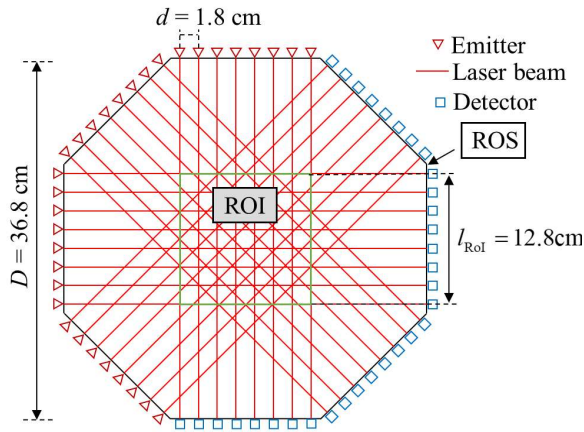

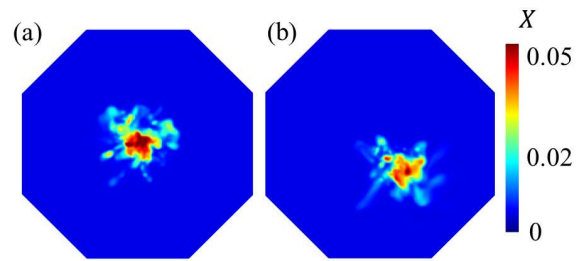

Fig. 6. Simulated phantoms of 2D distributions of $H_2O$ concentration in (a) scenario 1 (b) scenario 2, respectively. Phantoms (a) and (b) are corresponding to the 15th frames in Media 1 and Media 2, respectively.

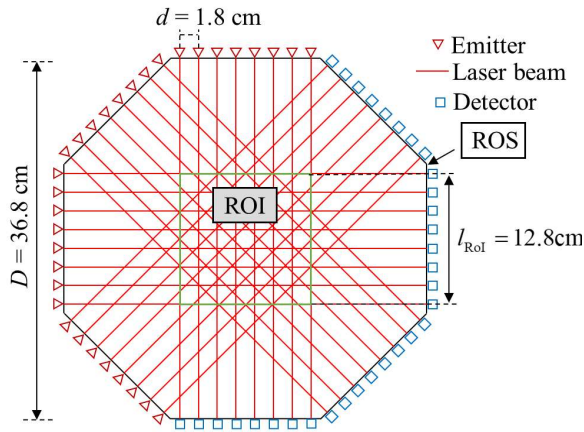

Fig. 5. Optical layout of the CST sensor.

and the true images [29]:

$$\text{IE} = \frac{1}{N} \sum_{j=1}^{N} \frac{\sqrt{\left(X_j^{\text{rec}} - X_j^{\text{true}}\right)^2}}{X_j^{\text{true}}}, \qquad (15)$$

where $X_j^{\text{rec}}$ and $X_j^{\text{true}}$ refer to the reconstructed and true $H_2O$ concentration at $j$-th pixel, respectively. Fig. 8 (a) and (b) shows the dependence of IE on $\gamma$ for the uniform-size and hybrid-size meshing schemes, respectively. Each IE value, displayed as a black dot in Fig. 8, was obtained by averaging the results from 50 repetitive reconstructions of a simulated phantom with the SNR of the line-of-sight TDLAS measurements set to 40 dB. For a given $\gamma$, the IEs that are displayed by the vertically distributed black dots in Fig. 8 were calculated from 5 consecutive sampled phantoms from each scenario, i.e. 11th to 15th frames in each Media. When $\gamma$ varies from $10^{-6}$ to 10 with 36 steps of logarithmic increment, the mean value of the IEs for each given $\gamma$ was interpolated as the red solid line shown in Fig. 8. The optimal $\gamma$ was selected where the minimum value of the mean IEs are obtained . The minimum mean IEs, i.e. IE = 0.35 and IE = 0.32, are obtained for the uniform-size and the hybrid-size meshing schemes given $\gamma = 4$ and $\gamma = 8 \times 10^{-4}$, respectively. Although similar IEs are obtained using both meshing schemes, the regularisation parameter $\gamma$ used in the uniform-size meshing is significantly larger than that used in the hybrid-size meshing. That is to say, the regularisation imposed on the uniform-size meshing strongly penalises the accuracy of the reconstructed image to trade off robustness [30]. As a result, the large regularisation term in (12) leads to disregard of the detailed flow characteristics and thus a poor-quality of the reconstruction despite the IE is acceptable. In addition, the uniform-size meshing scheme strongly depends on the regularisation parameter. As shown in Fig. 8 (a), a decreasing tendency of IE can be observed as $\gamma$ increases. In contrast, the hybrid-size meshing is less dependent on regularisation. The curve of mean IE shown in Fig. 8 (b) is mostly flat when $\gamma$ varies from $8 \times 10^{-4}$ to $2 \times 10^{-2}$, indicating the hybrid-size meshing scheme mitigates the ill-posedness from a deeper mathematical aspect, i.e. improvement of the sensing matrix, rather than the *a prior* information imposed on the target phantom. Therefore, it is more capable of imaging turbulence with more flow dynamic information including convection caused molecular movement and heat dissipation. For the ART and TV methods, the selection of the regularization goes through the same process as the above-mentioned for the TK methods.

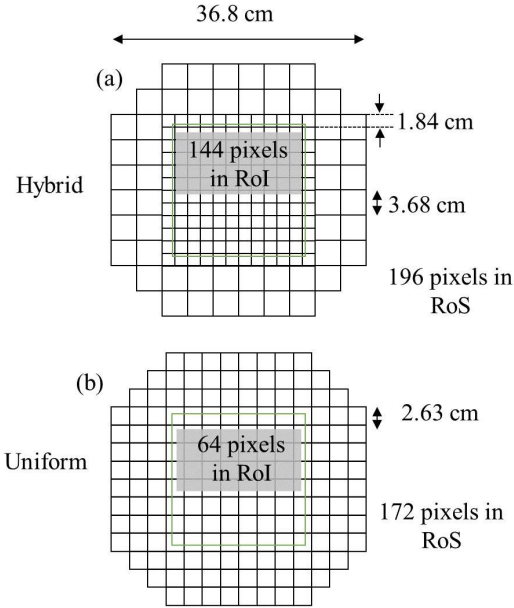

Fig. 7 Discretised RoS with (a) hybrid-size meshes and (b) uniform-size meshes.

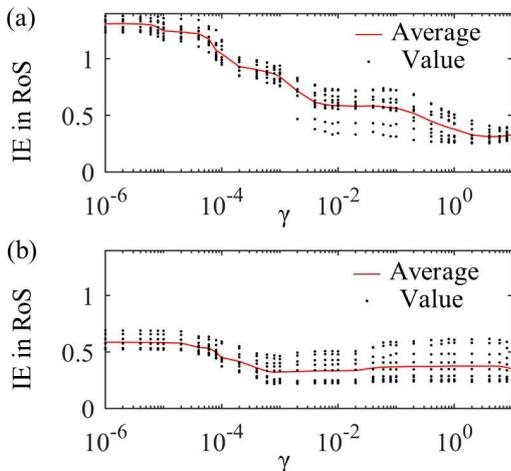

Fig. 8. Dependence of IE on $\gamma$ for the (a) uniform-size and (b) hybrid-size meshing schemes when reconstrting using TK.

Fig. 9 shows the reconstructed images of the two phantoms in Fig. 6 with the proposed hybrid-size and the typical uniform-size meshing schemes using the three different algorithms. Overall, the hybrid-size meshing scheme significantly outperforms the uniform-size meshing scheme in the following two aspects: (a) better accuracy with finer details of the inhomogeneity in the RoI; (b) fewer artefacts in the RoS. The former enables better retrieval of the profile and absolute concentration of the inhomogeneity, while the latter contributes to an accurate reconstruction of the background. Furthermore, the effectiveness of the hybrid-size meshing scheme was examined by reconstructions of the simulated $H_2O$ evaporation processes. The reconstructed images with the hybrid-size meshing scheme, using TK as an example shown in Media 1 and Media 2, can reveal more structural information of the cross-sectional variation of $H_2O$ concentration. For the whole $H_2O$ evaporation process, the absolute concentration of the inhomogeneity retrieved by the hybrid-size meshing scheme is also closer to the truth. It is also worth mentioning the hybrid-size meshes has 24 more pixels in the RoS. Given the same number of laser beams, the slightly larger number of pixels actually results in more undetermined solutions compared with the uniform-size meshing scheme. However, the improvement achieved by the hybrid-size meshing on the entire image quality significantly outweighs this defect.

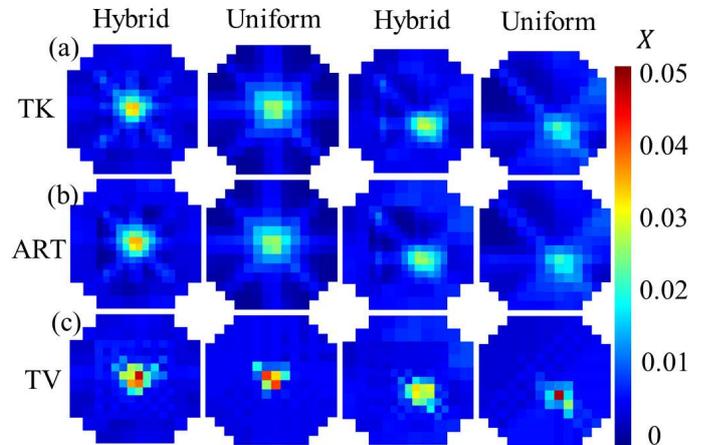

Fig.9 Reconstructions of the phantoms in Fig. 6 (a)using TK method, (b) using ART method, and (c)using TV method with different meshing schemes at 40 dB SNR, respectively.

Moreover, the performances of the two meshing schemes were quantified using simulated CST measurements contaminated with different levels of noise. Fig. 10 (a-c) and (d-f) show the IEs of reconstructed distributions of $H_2O$ concentration in the RoI and RoS at different SNRs using the three reconstruction algorithms, respectively. The value of IE at a given SNR is the average of IEs obtained from reconstruction of the above-mentioned 5 consecutive phantoms for each scenario. The hybrid-size meshing scheme gives persistently lower IEs in both the RoI and RoS using all the three tomographic algorithms, indicating higher accuracy for pixel-wise concentration. The improvement on image quality by the hybrid-size meshing is less significant at very low measurement SNR but becomes more significant as SNR increases. For practical CST measurements with SNRs higher than 35 dB [7], the biggest improvement on IE with the hybrid-size meshing is using ART, which gives more than 25% improvement on both RoI and RoS. For TK and TV algorithms, the hybrid-size meshing scheme shows more than 7% and 5% improvement on IE in RoI and RoS compared those with the uniform-size meshing, respectively.

## IV. EXPERIMENTAL VALIDATION

A lab-scale experiment was conducted to validate the proposed hybrid-size meshing scheme. The CST sensor was built with the same optical layout as depicted in Fig. 5. Distributed feedback laser diode (NLK1E5GAAA, NTT Electronics) was used to emit laser at the transition $v = 7185.6$ $cm^{-1}$. The laser current was driven with a sinusoidal scan at $f_s =$ 200 Hz superimposed by a sinusoidal modulation at $f_m = 40$ kHz. The pigtailed fibre from the laser diode was split by a 32-fibre splitter to deliver the 32 laser beams in the tomographic sensor. Each fibre-coupled laser was collimated and then detected by an InGaAs photodetector (G12182-010K, Hamamatsu). Then, the 32-way transmitted signals were digitised by a data acquisition platform, i.e. RedPitaya [31], at 3.9 Mega Samples/second. Digital lock-in modules were used to extract the Wavelength Modulation Spectroscopy (WMS) first and second harmonics, i.e. $1f$ and $2f$, from the transmission signals. Finally, the calibration-free $1f$ normalised $2f$ signal was fitted to calculate each of the 32 path integrated absorbances [32, 33].

In the experiment, each 2D distribution of $H_2O$ concentration was generated by $H_2O$ evaporation from a container filled with water at 60 degree Celsius. In the first case shown in Fig. 11 (a), a larger container with 14 cm diameter was placed at the centre of the RoS. As shown in Fig. 11 (b), the second phantom was generated by placing a smaller container with 8 cm diameter bottom right to the centre. The cross section of CST measurement was set 1 cm above the water surface. The ART algorithm described in Section II.C was used for image reconstruction. As detailed in Section III.B, the optimal regularisation parameter $\gamma$ was set as 4 and $8 \times 10^{-4}$ for the uniform-size and hybrid-size meshing schemes, respectively. The temperature in the RoS was assumed to be uniform and was read from a thermal sensor as 25 degree Celsius for both cases. As shown in Figs. 11 (c) and (d), the reconstructed $H_2O$ distributions using TK algorithm as an example, with the hybrid-size meshing scheme have more details in the RoI and fewer artefacts in the background than those using the uniform-size meshing scheme. For case 1, Fig. 11 (c) shows the convincing distribution of water vapour during transpiration with the finer resolved pixels in the RoI. However, the reconstruction using the uniform-size meshing shown in Fig. 11 (e) is hard to reveal any small-size $H_2O$ inhomogeneities beside the central inhomogeneity that are potentially introduced by the transpiration. For case 2, both reconstructed images can reveal the smaller size of the inhomogeneity with correct localisation in the RoS. With a better resolution in RoI shown in Fig. 11 (d), the reconstruction using the hybrid-size meshing is capable of indicating the gradient of the $H_2O$ concentration around the container. However, the reconstructed inhomogeneity in Fig. 11 (f) not only suffers from a much lower peak value of the $H_2O$ concentration, but also a severely blurred inhomogeneity due to the imposed strong smoothness regularisation. Therefore, the experimental results show the hybrid-size meshing scheme for CST is better at characterising the target flow fields with improved quality of reconstructed images.

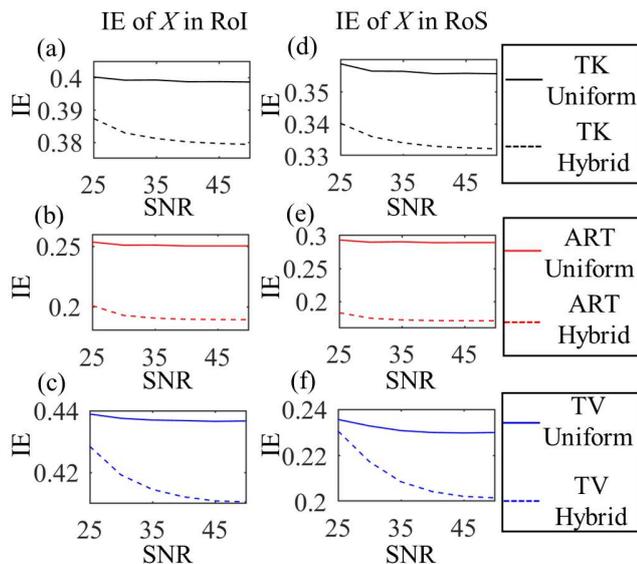

Fig. 10. Comparison of image errors in the reconstructed (a-c) RoI and (d-f) RoS with uniform-size and hybrid-size meshing schemes using the three algorithms at different measurement SNRs

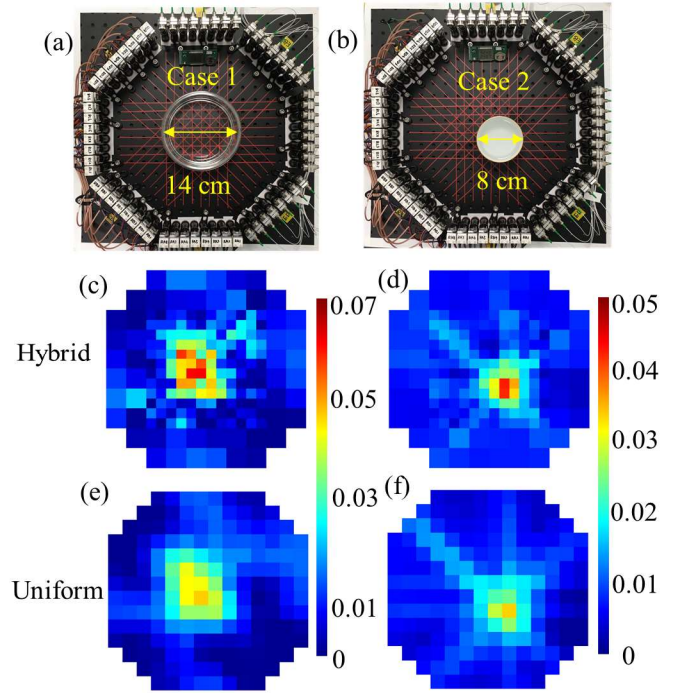

Fig. 11. Experimental reconstruction of the 2D $H_2O$ distributions. (a, b) Two experimental phantoms and the reconstructed distributions of $H_2O$ concentration using the (c, d) hybrid-size and (e, f) uniform-size meshing schemes.

## V. Conclusion

A new target-dependent CST with hybrid-size meshing scheme was proposed in this paper to improve the quality of images reconstruction. According to the location of target flow fields and given beam arrangement, the proposed method assigns smaller-size and larger-size pixels in and out of the RoI, respectively. This manipulation alleviates the ill condition of the sensing matrix in the CST inverse problem, resulting in better accuracy of tomographic reconstruction. Compared with the traditional uniform-size meshing scheme, the proposed scheme also achieves better resolution in the RoI, enabling finer characterisation of the target flow field.

Numerical simulations were carried out by reconstructing phantoms of 2D distributions of $H_2O$ concentration using both the hybrid-size and uniform-size meshing schemes. Using a variety of computational tomographic algorithms, i.e. TK, ART, and TV, the simulation results show the hybrid-size meshing scheme allows better accuracy with finer details in the RoI and fewer artefacts in the RoS. Furthermore, experimental validation was carried out by reconstructing transpiration-introduced inhomogeneity of $H_2O$ concentration. The experimental results show the correct localisation of the inhomogeneity with the ability to reveal convincing details of the target flows.

## Acknowledgement

The authors would also like to thank for Dr. Nick Polydorides and Prof. Hugh McCann for their useful suggestions.